# Definition of Exoplanets and Brown Dwarfs

Chapter 29 of the Handbook of Exoplanets

**Jean Schneider**

**Contents**

Introduction
What Is a Planet? .
From the Heaven of Concepts to the Hell of Observations
A Future Perspective
Conclusion
References

**Abstract**
This chapter reviews the definition of exoplanets and of brown dwarfs. Emphasis is given to the separation of these two populations. A traditional view is to declare « planet » objects with a mass < 13 $M_{Jup}$ and « brown dwarf » objects with a mass > 13 $M_{Jup}$ . By analogy with Solar System planets, a better definition is to call « planets » objects formed by accretion of dust and planetesimals in a disk. An by extension of the primitive introduction of the word « brown dwarf » for failed stars by Jill Tarter, this term must be reserved to objects formed by gravitational collapse in a molecular gas cloud. The two definitions do not coincide since a « brown dwarf » can have a mass down to about 6 Jupiter mass. And there is no physical reason to assert that a 20 Jupiter mass object has not been formed by accretion. From there, the diffuculty is to decide if an object of say 20 Jupiter mass is formed by dust accretion or by gravitational collapse. A future observational test to solve this difficulty is presented.

---

J. Schneider
LUTh, UMR 8102, Observatoire de Paris, 5 place Jules Janssen, F-92195 Meudon Cedex, France
e-mail: Jean.Schneider@obspm.fr

**Introduction**

Thousands of substellar objects (that is, brown dwarfs and planets) have been detected since 1989 (see, for instance, the following web sites: https://exoplanet.eu and
 https://exoplanetarchive.ipac.caltech.edu ).

Their masses run from 0.02 times the mass of the Earth (PSR 1257 C 20 b) to about 63 Jupiter masses (CoRoT-15 b), about five orders of magnitude in mass. More than 2700 confirmed planets have been detected by the photometric transit technique (e.g., see Chap. 4, "Discovery of the First Transiting Planets" by Dunham in this Handbook of Exoplanets), which led to an accurate determination of their size. Their radii run from 0.32 Earth radius (Kepler-37 b) to 2.1 Jupiter radius (HAT-P-67 b), about two orders of magnitude in size. The corresponding mass-radius diagram is represented in Fig. 1.

One could betempted to think that the more massive the object is, the larger it is in size and that there is some limit in mass and/or radius that distinguishes planets from everything else, even if this mass limit is below the well accepted substellar borderline at 72 Jupiter masses (minimum mass required for stable nuclear fusion of hydrogen in the interiors of solar metallicity stars). The objects between planets and very low-mass stars have been named brown dwarfs since the paper by Jill Tarter in 1973.

But Fig. 1 shows that beyond 1.2 Jupiter mass, there is a degeneracy in the objects' radii. One is then facing two problems: terminology (what is a planet? What is a brown dwarf?) and classification (how to decide if a given object is a planet or a brown dwarf according to a given definition?). Let us discuss these twao aspects.

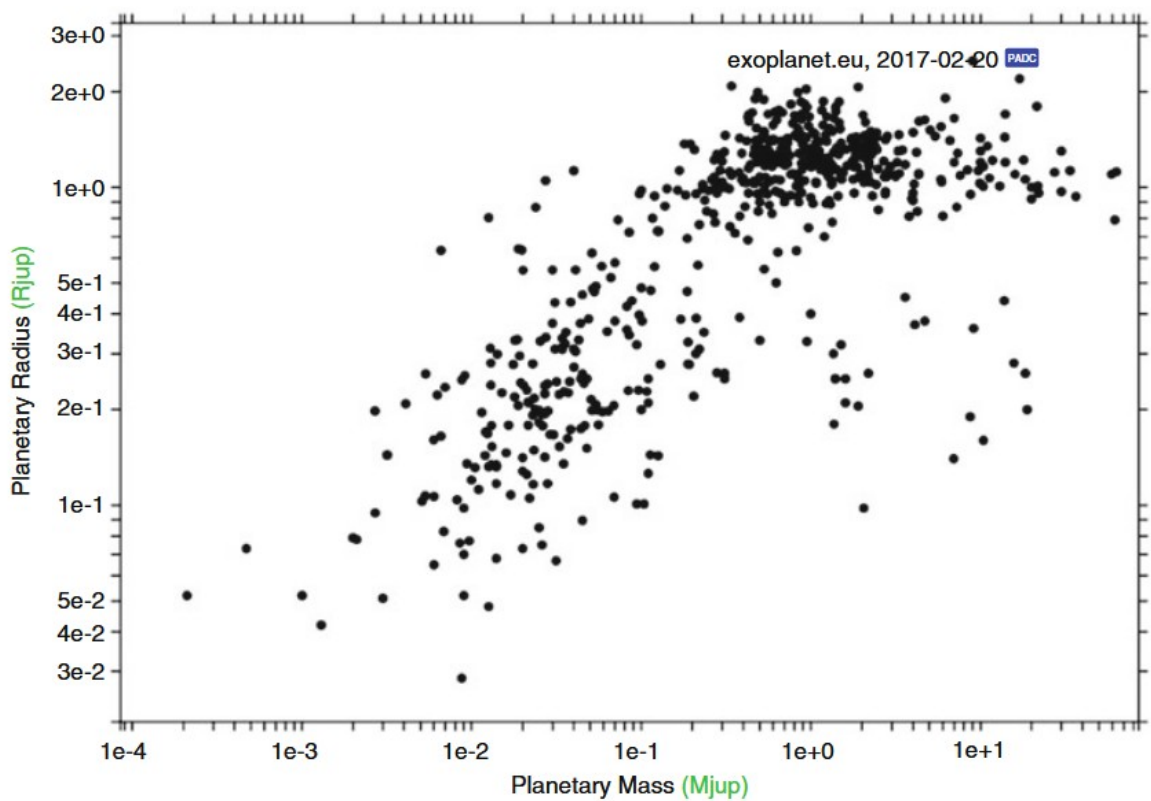

**Fig. 1** Mass radius relation for substellar objects (24 February 2017) from exoplanet.eu

**What Is a Planet?**

The debate, strongly motivated by the discovery of the first planets orbiting stars at the end of the last century (see Chap. 1, "The Discovery of the First Exoplanets") and the finding of planetary-mass objects in isolation at the beginning of this century (see Chap. 22, "Brown Dwarfs and Free-Floating Planets in Young Stellar Clusters") and encouraged by several authors (see, e.g., Baraffe et al. 2010; Schneider et al. 2011; Hatzes and Rauer 2015; Schlaufman 2018), is still ongoing and will not be closed by the present contribution.

Names are arbitrary conventions, but the natural trend is to classify objects using sufficiently elaborated concepts. Derived from the solar system analogy, one is tempted to call "exoplanets" (in short planets) to small bodies that are orbiting around other stars and were formed by condensation in a circumstellar dust disk. A first question is of course how small or massive the body has to be for being a planet. The problem here is that there actually exist small bodies orbiting stars that probably did not form like planets but like stars and brown dwarfs; this is from the collapse and fragmentation of a (possibly dusty) gas cloud.

**From the Heaven of Concepts to the Hell of Observations**

Formation provides a clear conceptual discrimination between planets and brown dwarfs (keeping in mind that it is a convention). But it is based on a criterion involving an inobservable concept, namely, the formation scenario, because we do not have brown dwarfs and planets birth movie at hand (see also Chap. 21, "Brown Dwarf Formation: Theory" by Whitworth and the section on "Formation and Evolution of Planets and Planetary Systems" in this Handbook of Exoplanets).

One can only rely on actual observables. Standard basic observables are the object mass, radius, and temperature. An ideal situation would be that, at least for one of these observables, there exist two domains $D_{planet}$ and $D_{brown\ dwarf}$ of values which do not intersect. It is unfortunately not the case since there are, according to formation models, objects formed by condensation of dust (i.e., planets according to the proposed convention) (smaller or larger, heavier or lighter, cooler or hotter) than objects formed by collapse (i.e., brown dwarfs). Even worse, there are a few pulsar companions with masses lower than 30 Jupiter mass. They are probably the relic of stellar companions eroded by the pulsar strong wind (Ray and Loeb 2015). One can argue that as such they are not planets nor brown dwarfs, their formation process being very different. But one cannot exclude that such erosion mechanism happened also for low-mass companions of main sequence stars with strong winds (see e.g., Sanz-Forcada et al. 2010). Consequently, the choice between "planet" and "brown dwarf" classification for a given object can only be arbitrary. The choice made by the Extrasolar Planets Encyclopaedia at exoplanet.eu is to consider all objects below 60 Jupiter mass as "planets," based on the results by Hatzes and Rauer (2015). Hatzes and Rauer argument is that the mass-radius and the mass-density relations present a well-defined increasing trend for objects more massive than Saturn (giant planetary régime) up to a certain mass value where the slope of the trend changes dramatically. This happens at 60 Jupiter mass (Fig. 2). Unfortunately, the number (statistics) of objects in the 30–60 Jupiter mass region is very low (the so-called brown dwarf desert); for producing Fig. 2, Hatzes and Rauer (2015) used only transiting planets and brown dwarfs and did not correct the relation shown in Fig. 2 from the mass distribution (or mass histograms) at different mass intervals. Earlier data suggested a dip around 40 Jupiter mass (Sahlman et al. 2011, Udry et al. 2010 – see also Figs. 3 and 4) in the mass histogram. More statistics will become available in the near future thanks to radial velocity surveys from ground facilities and astrometric data from Gaia that will allow to investigate whether the feature around 40 Jupiter mass in the mass-radius diagram exists or not.

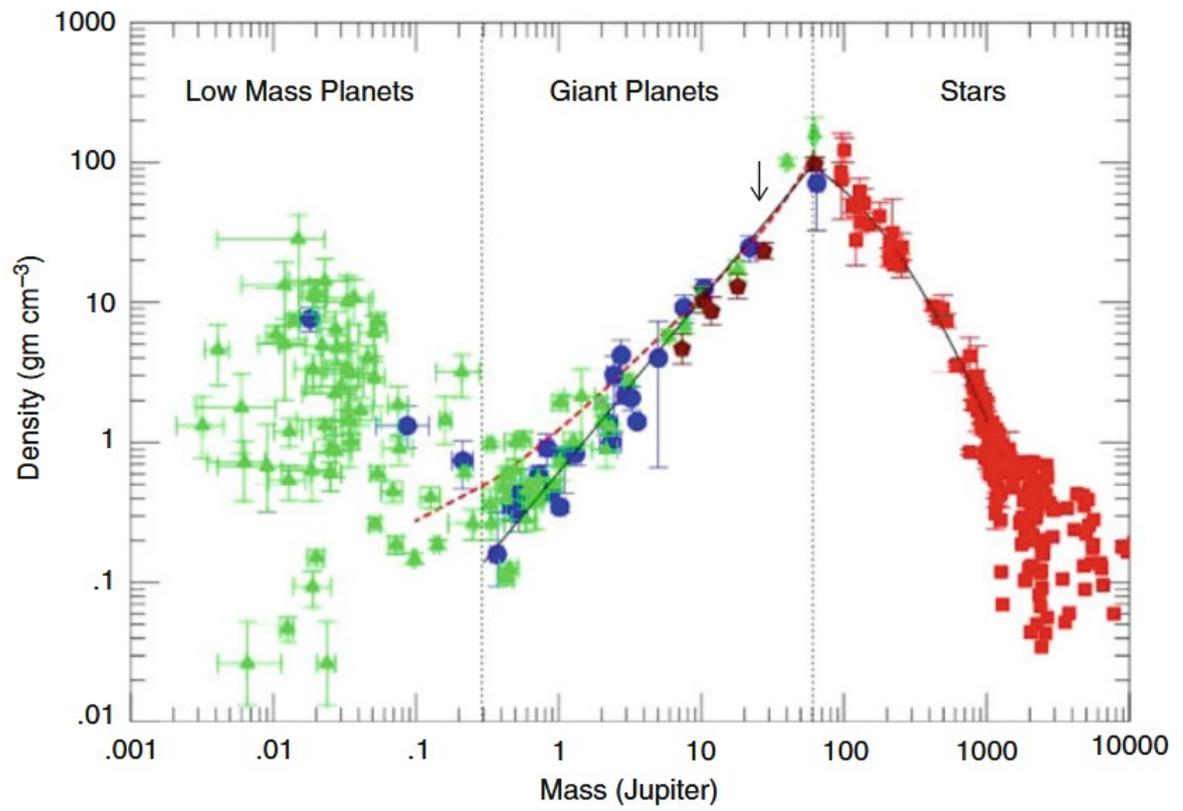

**Fig. 2** Empirical mass-density relation (Hatzes and Rauer 2015)

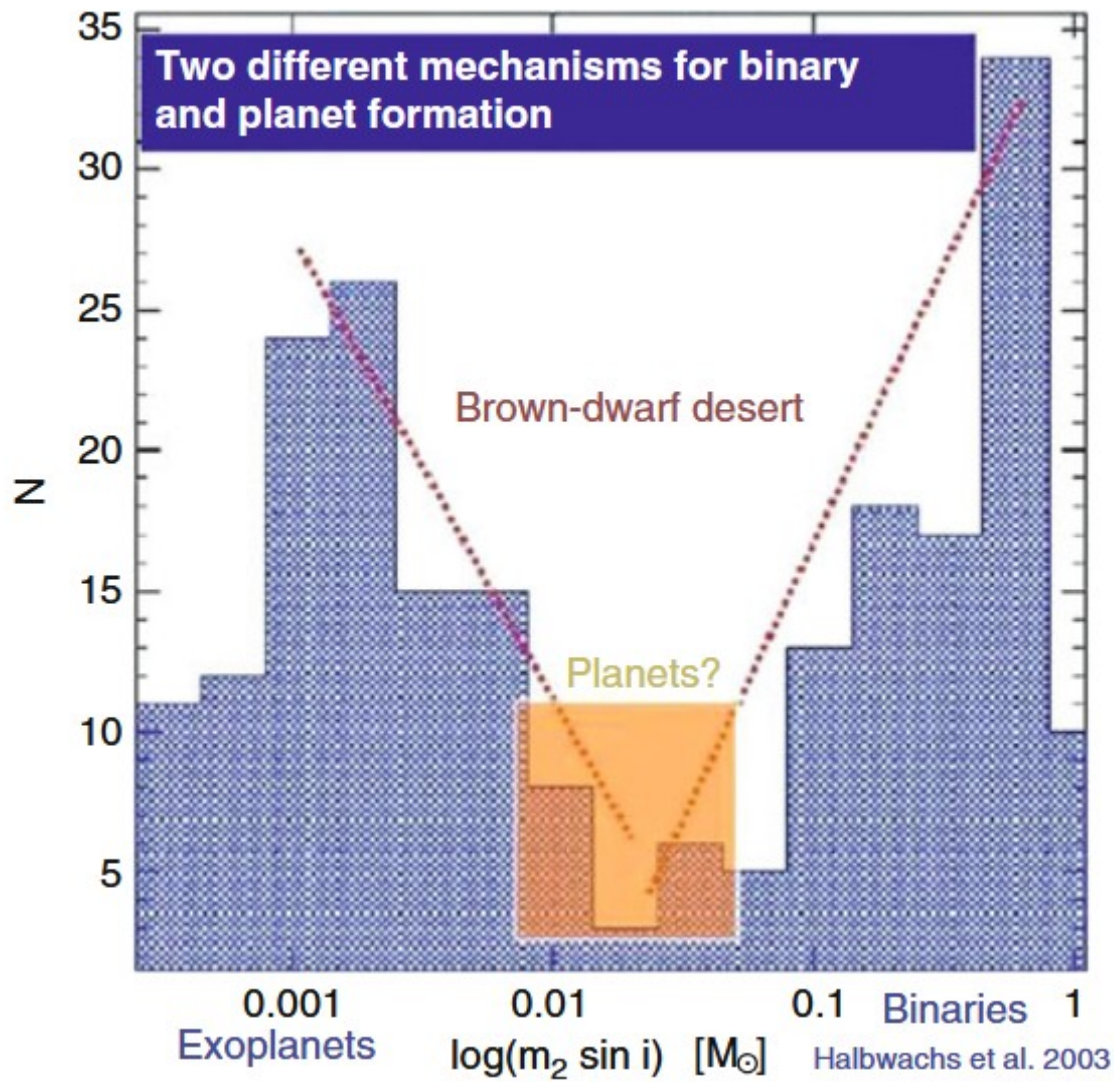

**Fig. 3** Mass histogram for low-mass objects (Udry et al. 2010)

A future improvement to separate the planet and brown dwarf populations will likely be possible from advanced observables, like the spectral type and species composition. They will help to constrain the formation mechanism of the object Definition of Exoplanets and Brown Dwarfs (accretion in a dust disk or collapse and fragmentation from a gas cloud). At least one conclusion is clear: the former criterion to discriminate planets and brown dwarfs based on the mass boundary at 13 Jupiter masses, corresponding to the triggering of nuclear burning of deuterium at the interiors of the small bodies, is not relevant since, theoretically speaking, objects can be grown by dust accretion from a circumstellar disk (planet-like formation) and acquire a final mass larger than 13 Jupiter masses.

There is a second, more factual problem: the value of some observables, particularly the mass, can be very uncertain. This is especially the case for companion objects detected by direct imaging where the mass cannot be inferred from radial velocity measurements but only from spectra, photometry, and models. A typical example is the object 2M1207b (Chauvin et al. 2004), which is located at a distance of >55 AU from its parent brown dwarf and has a mass of 4 ±1 Jupiter mass. Indeed, in these cases the planet-star or planet-brown dwarf separation is so wide that the semi-amplitude $K = \sqrt{GM_*/a_{pl}}$ of the parent object radial velocity variation induced by the planet motion is too low to be measurable with current technology. Even more, when the mass determination is as precise as a few percent (in case of radial velocity or astrometric measurements), one faces the absurd situation of a sharp mass limit. For example, how should we classify objects like CoRoT-15 b that has a mass at the borderline, $M = 63.3 \pm 4\, M_{Jup}$?

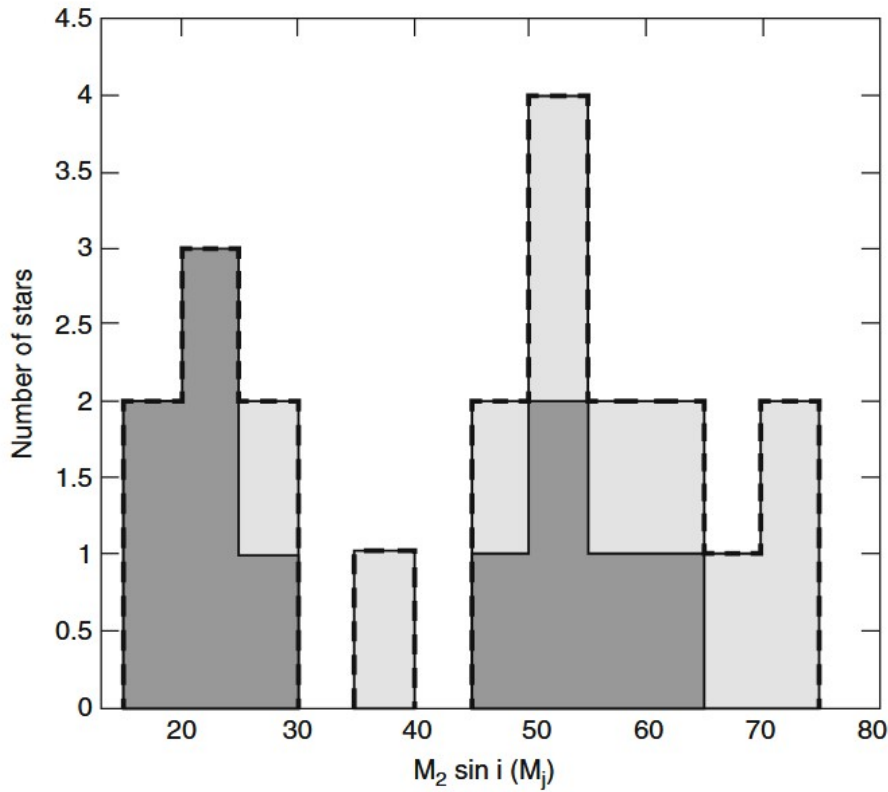

**Fig. 4** Low-mass object histogram in the 20–75 Jupiter mass region (Sahlman et al. 2011)

An additional problem, which will not be addressed here, is the existence of the "interstellar wanderers," i.e., planets that are free-floating (they do not orbit around any more massive object) either because they were formed in an isolated way or because they were expelled by dynamical interactions within multiple planetary systems during the early stages of the formation of the system. For more information regarding this population, see the section "Between Planets and Stars: Substellar Objects » in this Handbook of Exoplanets.

**A Future Perspective**
There is, nevertheless, a future hope to distinguish planets from brown dwarfs by inspecting the internal structure thanks to spectrophotometric seismology with future ground-based, lunar, or spatial very large telescopes. Indeed the core of a 20–20 Jupiter mass object should not have the same equation of state if it is formed by planetesimal accretion or by gravitational collapse of a gas.

**Conclusion**
Assuming that the definition of planets and brown dwarfs is adopted according to their formation mechanism, to separate the two populations is not an easy task. Any catalog contains necessarily a mixture of both populations. Since catalogs are useful not only to list the objects properties but also to make statistics regarding these properties, this author recommends to take low constrains (a mass limit as high as 60 Jupiter mass is adopted at exoplanet.eu) on the properties used to define a sample, in order not to miss interesting objects. Modern software used to read electronic catalogs allows to eliminate easily objects from a catalog that do not fulfil the criteria of each user, who is free to impose his/her own criteria.